# Quantum Randomness and Nondeterminism

E. Knill
knill@lanl.gov, B265, Los Alamos National Laboratory, Los Alamos, NM 87545; and Institute for Theoretical Physics, University of California, Santa Barbara, CA 93106-4030.

## Abstract

Does the notion of a quantum randomized or nondeterministic algorithm make sense, and if so, does quantum randomness or nondeterminism add power? Although reasonable quantum random sources do not add computational power, the discussion of quantum randomness naturally leads to several definitions of the complexity of quantum states. Unlike classical string complexity, both deterministic and nondeterministic quantum state complexities are interesting. A notion of *total quantum nondeterminism* is introduced for decision problems. This notion may be a proper extension of classical nondeterminism.

## 1 Introduction

Quantum algorithms are based on applying local unitary operations to coherent superpositions of states. For discussions of the basic principles see the rapidly growing literature on the subject, e.g. [10, 9, 7, 3, 1] and many papers on http://xxx.lanl.gov/ in quant-ph. It is well known that quantum algorithms can simulate any classical probabilistic algorithm by exploiting quantum coin flips. Because of the ability to factor numbers in polynomial time using quantum computation [7], it is widely believed that quantum computation is strictly more powerful than classical probabilistic computation. However, the relationship between nondeterministic polynomial computation and deterministic polynomial quantum computation is far from established. In this note, we discuss notions of quantum randomness, quantum state complexity and proper extensions of nondeterminism to quantum computation.

## 2 Preliminaries

A straightforward model of quantum computation is that of a QRAM, which is a classical RAM with access to quantum registers. The quantum registers can be prepared in the (classical) initial state $|\mathbf{0}\rangle$ for the finite bit string $\mathbf{0}$, manipulated by a set of primitive one and two qubit unitary operators and measured in the classical basis. The classical basis is labeled by bit strings of the register's length. A QRAM can be programmed just like a RAM, which means that an algorithm is specified by a program [6]. The program can be provided by a classical bit string to a universal QRAM. Quantum information can be provided by supplying an initial quantum state in a quantum register. An algorithm for a QRAM is therefore specified by a classical program and (if desired) an initial quantum state.

The complexity of quantum states is explored in [4, 5]. There it is established that almost all quantum states on $n$ qubits require exponentially many local operations to approximate to within any distance better than random, both for the inner product norm and the total variation distance (of measurement distributions). This is a consequence of the observation that if $C$ is a subset of the unit Hilbert sphere on $n$ qubits such that every vector in the sphere is within distance $\epsilon < \sqrt{2}$ of a member of $C$, then the cardinality of $C$ is doubly exponential in $n$. As a result of these observations, only a very small fraction of the state space of reasonable numbers of qubits can be explored by any feasible computation. Since it is generally believed that in the real world, large scale coherence disappears on typically fairly short time scales, it would also appear that difficult to prepare states do not exist in nature. Could we gain computational power by having access to sources of such states?

## 3 Quantum Randomness

A *quantum random state* on $n$ qubits is a state picked randomly according to the uniform distribution on the Hilbert sphere in the $2^n$ dimensional complex Hilbert space generated by the qubits. Because of the discussion in the previous section, it would seem that such a source cannot be easily found or designed. However, because no additional information about the state is available, it can be shown that its effect on a quantum computation can be simulated by $n$ classical coin flips. To see this, note that the effect of the computation is completely described by considering the density matrix induced on the $n$-qubits by the random state. This density matrix is given by a scalar multiple of the identity, which is the same as that describing the outcome of $n$ classical coin flips.

To attempt to strengthen the definition of a quantum random source, one might consider adding the ability of repeatedly accessing the same random state. Thus a quantum random state $|\psi\rangle$ is made available in as many copies as needed by a suitable oracle. Note that unless we know how to generate $|\psi\rangle$ by unitary transformations from a reproducible initial state, we cannot make such copies from a single instance. An algorithm can make use of as many copies as needed. Can the effect of such a random source be effectively simulated? Let $\rho$ be the density matrix associated with $k$ copies of a random state $|\psi\rangle$. Let $|\psi\rangle$ be $d$ dimensional. The interesting case is where $d$ is exponentially large. The $k$-fold tensor product of $|\psi\rangle$ lives in a space whose basis can be labeled by all sequences of numbers between 1 and $d$. By explicitly integrating $|\psi\rangle^k\langle\psi|^k$, it can be seen that $\rho$ is block diagonal, with each block supported on the set of states labeled by sequences obtained by permuting a fixed sequence $s$. The blocks are rank one, given by $|b\rangle\langle b|$, where $|b\rangle$ is a uniform superposition of $|s'\rangle$'s with $s'$ a permutation of $s$. The trace of a block, i.e. the probability of $|b\rangle$, is an integral of products of even powers of coordinates over the $r$-dimensional sphere. On way to simulate the source requires generating a sequence with probability distribution determined by the multiset of numbers it contains, then coherently producing all permutation equivalent sequences. Problem: Give an effective method for implementing this procedure.

## 4 Quantum State Complexities

Although the obvious definitions of quantum random sources do not lead to additional computational power, the fact that most states are difficult to approximate from any given initial state leads to the problem of understanding quantum state complexities. It turns out that notions of quantum state complexity are substantially richer than the classical version. Consider the general problem of converting one quantum state $|\psi_i\rangle$ to another $|\psi_f\rangle$ by means of a quantum computer, where for simplicity we assume that $|\psi_i\rangle$ and $|\psi_f\rangle$ are supported on the same number of qubits, i.e. have the same length.

The most general method to convert $|\psi_i\rangle$ to $|\psi_f\rangle$ involves supplying a classical program and quantum information to a universal QRAM with $|\psi_i\rangle$ in the input register. In the exact model, the QRAM should halt with $|\psi_f\rangle$ in the output register. In the approximate model it should halt with $|\psi_f'\rangle$ in the output register, where $|||\psi_f'\rangle - |\psi_f\rangle|| < \epsilon$ for a given $\epsilon > 0$ which is considered part of the classical input. For simplicity, the output register is considered to be separate from the input register and the input register can be modified.

There are several resources that are of potential interest. They are time, space, program length (with and without quantum information), nondeterminism, and state preparation/measurement[1]. In all cases, we are interested in the resource requirements as a function of the length of $|\psi_i\rangle$ and the inverse of the approximation parameter $\epsilon$.

**Time.** In the classical case, the time resource is not interesting unless other resources are constrained. The conversion from one string to another can always be accomplished in linear time. In the quantum case, the results of [4] imply that nearly all states require exponentially many steps to approximate starting from $|\mathbf{0}\rangle$.

**Space.** One can consider both classical and quantum space requirements. By the results on completeness of the one and two qubit unitary operations in the unitary group, the computation can always be arranged to use no qubits other than those in the input and output registers. As a result, no space is required beyond that occupied by the program and the quantum inputs and outputs. Thus, except when considering trade-offs, the interesting space related resources are program length and state preparation/measurement.

**Program length.** This is the conditional Kolmogorov complexity. If the quantum information provided with the program is unrestricted, then a linear size program suffices to perform the state conversion, simply by providing the final state as part of the program. If the classical component of the program is unrestricted, then the conversion can be accomplished without quantum input. However, the classical component may be exponentially large. Thus there are two interesting situations that may be considered. In the first, the resource is the sum of the lengths of the classical program and the quantum information provided with the program. As an alternative, it may be reasonable to take the logarithm of the length of the classical program before adding the length of the quantum information. In the second situation, the resource is the classical program length and no quantum

---
[1] Note that state preparation can be viewed as measurement followed by a unitary operation depending on the outcome of the measurement.

information can be provided with the program.

**Nondeterminism.** For classical string conversion problems, nondeterminism trivializes resource needs. It suffices to nondeterministically generate each character. The input state does not contribute. Nondeterminism in a quantum computation can arise whenever measurements are performed as part of the calculations. We consider a state to be obtained nondeterministically, if it is the state associated with one of the measurement outcomes with non-zero probability. Using a nondeterministic method for producing a state does not guarantee that the state is actually obtained, but does ensure that we know if it has been obtained. The computation must be error-free to guarantee this. This type of nondeterminism is at least as powerful as the classical version. However, it does not trivialize the resource needs because the ensemble of states accessible by any fixed algorithm is at most exponential in the total number of qubits measured. Thus, for polynomially bounded computations the total number of nondeterministically accessible states is still the exponential of a polynomial, while a doubly exponential number of states is required to non-trivially approximate all states of a given number of qubits. We wish to determine whether and to what extent nondeterminism helps to reduce other resource requirements. One can attempt to quantify the degree of nondeterminism by the logarithm of the probability of observing the correct final state.

An interesting question is the following: Given a nondeterministic method for generating a state, what is the complexity of generating it deterministically?

**State preparation/measurement.** This resource is of interest primarily in the case where the QRAM is restricted to perform reversible logic and the program is not retained. See [2] for an explanation and analysis of the classical case.

In summary, in the quantum case there are many interesting state conversion complexities, depending on the resource considered, time, program length, or state preparation, whether or not the program can contain quantum information and whether or not the algorithm is constrained to be deterministic.

## 5 Quantum Nondeterminism

We have already introduced a notion of nondeterministic resource requirements for converting quantum states. This notion generalizes classical nondeterminism. Whether nondeterminism introduced by measurement can be simulated by nondeterminism in the classical choices made by the classical program supplied to the QRAM needs to be determined. Thus it may be the case that at least in the error-free model of quantum computation measurement introduces a genuinely new source of nondeterminism. In this section we consider a more traditional way of defining nondetermism for decision problems and how it may be strengthened for quantum computation.

The obvious notion of nondeterminism for quantum algorithms uses a nondeterministic input just as in classical computation. Using the language of relations, $\exists y R(x, y)$ is in $NQT(n)$ if there exists a quantum algorithm such that for $|x| \leq n$ and $|y| \leq O(T(n))$, it returns $R(x,y)$ in $O(T(n))$ steps with high probability. Here $|x|$ denotes the number of qubits of the input $x$. The problem is in $QT(n)$ if a quantum algorithm can determine on input $x$ in time $O(T(|x|))$ whether $\exists y R(x,y)$, with high probability.

Because nondeterminism is introduced by a classical choice, the notion of nondeterminism defined above is essentially classical. Is there anything to be gained by allowing $y$ in $R(x,y)$ to range over all quantum states? A quantum algorithm on quantum inputs $x$ and $y$ defines a probabilistic relationship $R(x,y)$ by the distribution induced on the ouput bit. One possible way of defining what it means for a quantum algorithm to solve $\exists y R(x,y)$ is to require that its output bit after measurement is near[2] $|1\rangle$ if for some $y$, $R(x,y)$ holds with probability $> 3/4$, and it is near $|0\rangle$ if for all $y$ $R(x,y)$ holds with probability $< 1/4$. In the other cases, the output is unconstrained. For this definition to be interesting, it should be the case that it is not sensitive to the exact choice of threshold probabilities. For now we just observe that the algorithm with nondeterministic input $y$ which simply computes $R(x,y)$ has the correct output distribution for some $y$.

A deterministic algorithm which solves $\exists y R(x,y)$ in the sense described works as follows: It searches through states $y$ in a sufficiently dense set by approximation using local operations, runs the algorithm for $R(x,y)$ sufficiently many times to establish the output distribution, producing output $|1\rangle$ if it is biased toward membership. If no bias is found for any of the sample $y$, the output is $|0\rangle$. Note that the input $x$ must be available as an independent copy for each trial in the search. It is sufficient for $x$ to be classical or specified by a program.

The algorithm for $\exists y R(x,y)$ of the previous paragraph needs exponential time just to generate any one of the states (except for a small fraction). Since a doubly exponential number of states needs to be tried, each output distribution has to be sampled sufficiently well (exponentially many times) to ensure success. The nondeterministic version of this algorithm which generates and tries a random state requires exponential time in general (to generate the state). Thus, this strong version of quantum nondeterminism, *total quantum nondeterminism*, seems to apply primarily to the exponential (uniform) complexity classes and beyond.

---

[2]E.g. the probability of seeing $|1\rangle$ is at least $3/4$.

It is not clear that the notion of total quantum nondeterminism is a good one, or even particularly useful. The property of an input $x$ that is defined is not determined. For some inputs, either of the possible outputs is allowed.

# 6 Conclusion

It has been shown that the obvious definition of quantum randomness does not add computational power. However, due to the difficulty of exploring much of the state space available to qubits, the notion of quantum state complexity is very rich, with several resources being of potential interest. Both program length (classical and quantum) and time are nontrivial. Several sources of nondeterminism have been discussed. Can nondeterminism introduced by measurement be simulated by classical choices made by the QRAM program without substantially increasing resource requirements? Although the classical version of nondeterminism is relevant to quantum computation, it may be interesting to consider total quantum nondeterminism, which replaces the classical choices with a quantum state.

# 7 Acknowledgements

This work was performed under the auspices of the U.S. Department of Energy under Contract No. W-7405-ENG-36 and supported in part by the National Science Foundation under Grant No. PHY94-07194.

# References

[1] D. Beckman, A. N. Chari, S. Devabhaktuni, and J. Preskill. Efficient networks for quantum factoring. quant-ph/9602016, 1996.

[2] C. H. Bennett, P. Gács, M. Li, P.M.B Vitányi, and W.H. Zurek. Thermodynamics of computation and information distance. In *Proceedings of the 25th ACM Symposium on the Theory of Computation*, pages 21–30, 1993.

[3] A. Ekert and R. Jozsa. Notes on quantum factoring. *Reviews of Modern Physics*, 1993.

[4] E. Knill. Approximation by quantum circuits. Technical Report LAUR-95-2225 and 68Q-95-29 at http://www.c3.lanl.gov/laces, Los Alamos National Laboratory, 1995.

[5] E. Knill. Bounds for approximation in total variation distance by quantum circuits. Technical Report LAUR-95-2724, Los Alamos National Laboratory, 1995.

[6] E. Knill. Conventions for quantum pseudocode. Technical Report LAUR-96-2724, Los Alamos National Laboratory, http://www.c3.lanl.gov/~knill, 1996.

[7] P. W. Shor. Algorithms for quantum computation: Discrete logarithms and factoring. In *Proceedings of the 35'th Annual Symposium on Foundations of Computer Science*, pages 124–134. IEEE Press, 1994.

[8] P. W. Shor. Fault-tolerant quantum computation. quant-ph/9605011, to appear in FOCS 1996, 1996.

[9] D. R. Simon. On the power of quantum computation. In *Proceedings of the 35'th Annual Symposium on Foundations of Computer Science*, pages 116–123. IEEE Press, 1994.

[10] A. Yao. Quantum circuit complexity. In *Proceedings of the 34th Annual Symposium on Foundations of Computer Science*, pages 352–360. IEEE Press, 1993.